# Deep Conditional Shape Models for 3D cardiac image segmentation

Athira J Jacob[1,2], Puneet Sharma[1] and Daniel Ruckert[2]

[1] Digital Technology and Innovation, Siemens Healthineers, Princeton, NJ, USA
[2] Chair for AI in Healthcare and Medicine, Klinikum rechts der Isar, Technical University of Munich, Germany
`athira.jacob@siemens-healthineers.com`

**Abstract.** Delineation of anatomical structures is often the first step of many medical image analysis workflows. While convolutional neural networks achieve high performance, these do not incorporate anatomical shape information. We introduce a novel segmentation algorithm that uses Deep Conditional Shape models (DCSMs) as a core component. Using deep implicit shape representations, the algorithm learns a modality-agnostic shape model that can generate the signed distance functions for any anatomy of interest. To fit the generated shape to the image, the shape model is conditioned on anatomic landmarks that can be automatically detected or provided by the user. Finally, we add a modality dependent, lightweight refinement network to capture any fine details not represented by the implicit function. The proposed DCSM framework is evaluated on the problem of cardiac left ventricle (LV) segmentation from multiple 3D modalities (contrast-enhanced CT, non-contrasted CT, 3D echocardiography-3DE). We demonstrate that the automatic DCSM outperforms the baseline for non-contrasted CT without the local refinement, and with the refinement for contrasted CT and 3DE, especially with significant improvement in the Hausdorff distance. The semi-automatic DCSM with user-input landmarks, while only trained on contrasted CT, achieves greater than 92% Dice for all modalities. Both automatic DCSM with refinement, and semi-automatic DCSM achieve equivalent or better performance compared to inter-user variability for these modalities.

## 1 Introduction

Delineation of anatomical structures is a fundamental task in medical image analyses and often forms the first step in many clinical quantification and diagnoses workflows. Deep learning (DL) has become the most widely used approach for cardiac image segmentation over the recent years[1]. While techniques such as fully convolutional networks demonstrate state-of-the-art segmentation with accuracy[2], they have some important weaknesses: First, unlike natural images, the domain of medical images face the issue of data and/or label scarcity[3]. Annotating medical images can be expensive and tedious, often requiring trained clinical experts. In addition, it is difficult to generalize domain knowledge across different modalities, even though the underlying anatomy might be the same. Utilizing information across various modalities becomes even more critical when some modalities inherently have more or higher fidelity data.

To address these challenges, we propose the Deep Conditional Shape models (DCSMs), an intuitive approach to anatomical segmentation using shape priors (Fig 1). Using deep implicit shape representations[4–7], a modality agnostic shape model is learnt that can represent the anatomy of interest via signed distance functions. To fit the generated shape to the image, the shape model is conditioned on anatomic landmarks that can be automatically generated or provided by the user. Finally, we add a modality dependent, lightweight refinement network to capture any fine details not represented by the implicit function. As a proof of concept, we evaluate DCSM on the problem of cardiac left ventricle (LV) segmentation from multiple 3D modalities (contrast-enhanced CT, non-contrast CT, 3D echocardiography – 3DE). The DCSM is trained on a large dataset of contrasted CT from 1474 patients and evaluated on all modalities. Compared to the baseline, the automatic DCSM obtains 3.6 to 28.9 mm decrease in average Hausdorff distances, and 0.2 to 7.4 point decrease in average relative volume errors, depending on the modality. In the case of non-contrast CT and 3DE, the semi-automatic DCSM obtains an increase of 13 and 6 points in average Dice scores and 35 and 9 mm decrease in average Hausdorff distances.

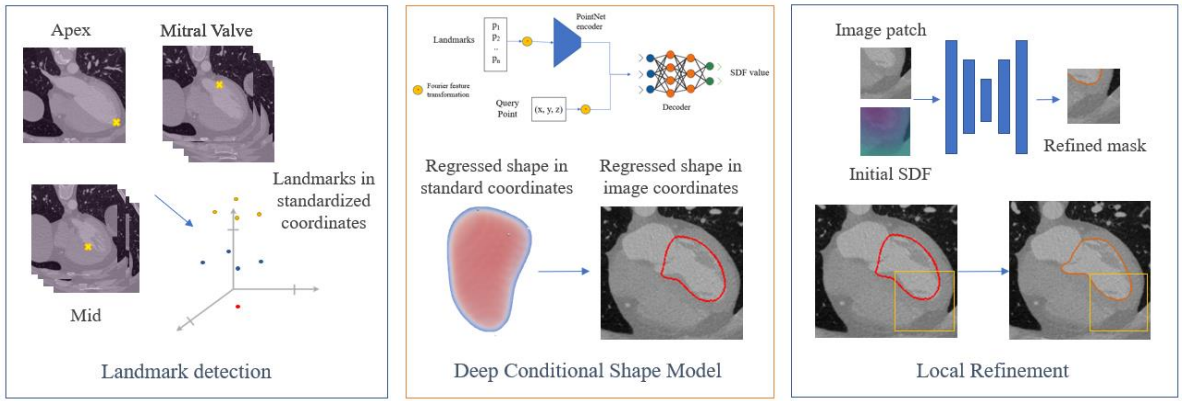

**Fig. 1**. Proposed method. 1) Landmark detection: 9 anatomically relevant reference points are detected from the image and converted to standard coordinate system. 2) Shape regression: A deep conditional shape model, trained on a dataset of LV shapes, outputs a LV segmentation conditional on the provided landmarks. 3) Local refinement: For modalities with image information, a lightweight image to image model is used to refine the initial detection using local information. SDF: Shape distance function

3## 2 Related Work

**Shape priors in medical image segmentation.** Shape priors offer a way to integrate anatomical information into segmentation tasks. Shape priors have been traditionally used in segmentation through statistical shape models [13]. Recent methods in deep learning have tried to learn joint image and mask representations from data to encode the shape implicitly[6] through additional neural networks, cost functions, variational learning [12] or adversarial training [13]. These methods require training modality-specific neural networks, or only modelling the shapes implicitly. Graph based methods [14] have also been explored for medical image segmentation. While these model shapes explicitly, training such models often require dense meshes with point-to-point correspondences, and can only model fixed topologies.

**Neural shape representations.** Recently, neural implicit functions have been used extensively as shape representations[5,6]. A neural network is fed with a latent code and a query point coordinate to predict a signed distance function or the binary occupancy at that location. Querying continuous points used in implicit function learning allows predicting in continuous space, creating robustness to a wide range of resolutions, and being more computationally efficient than voxel-based methods. It also allows for learning from partial data. Implicit shape functions have shown state of the art results in 3D shape generation[15], reconstruction[17, 18], hyper-resolution[19] etc.

**Multi-modality segmentation.** Transfer learning has been used to learn from modalities with large, annotated datasets and improve performance in a target domain with more scarce data/annotations through fine-tuning[22], multi-task learning [24] and adversarial training [26] . Unsupervised domain adaptation forms another category of methods, including image translation where images of a more scarce modality are generated from data and annotations in a source domain [26, 27] and few shot learning methods[28, 29] where a model is trained primarily in the source domain, and requires very few or no training examples from the target domain.

## 3 Proposed Method

An overview of the proposed model is given in Fig 1.

### 3.1 Shape regression: Conditional implicit shape model

We learn a signed distance function (SDF), which is a continuous function that represents the distance of each spatial point to the closest surface. The surface of the shape is represented by the zero level-set of the neural net. To learn the positioning of the anatomy in 3D image space, we condition the model on input landmarks. Given a set of standardized landmark points $P = \{p_i\}_{i=1}^{N} \in \mathbb{R}^3$ with N points, and a query point $x \in \mathbb{R}^3$, we learn a conditional $SDF$ $\phi$ such that, $SDF(x, P) = \phi(\gamma(x), \gamma(P))$, which gives the signed distance of point $x$ from the nearest LV surface (+ for outside and − for inside the structure). Here, $\gamma(P)$ is a Fourier feature-based mapping[30], defined as $\gamma(P) = [\cos(2\pi BP), \sin(2\pi BP)]^T$, where each entry in $B \in \mathbb{R}^{m \times 3}$ is sampled from gaussian $\mathcal{N}(0, \sigma^2)$. The use of Fourier feature mapping facilitates enhanced learning of the high frequency information, as compared to coordinate-based approach. We perform this mapping for both $x$ and $P$. The effect of number of input landmarks was studied and given in Appendix 1.1. The effect of the Fourier Mapping is shown in Appendix 1.2.

We train the continuous SDF model on a database of parametric LV meshes. We first preprocess the data to extract SDF samples for each training mesh. All training samples are standardized by affine alignment to a mean mesh $Q$ calculated from the training data and centered in a unit cube. We sample 100,000 spatial points from the unit cube along with their SDF values, with around 80% of the points lying near the surface.

For our experiments, we use a modified PointNet with residual connections[31] as the encoder architecture. We use 5 down-sampling blocks, with both hidden dimension and latent dimension of 128. The decoder consists of 5 blocks, with each block consisting of 1D convolutions, batch norm and ReLU layers, with a hidden size of 256. We use $L_1$ loss function and ADAM optimizer, with a learning rate of $1e^{-4}$. We set $m = 32$ and $\sigma = 1$ for the Fourier feature mapping. We also train the shape model with random, normally distributed perturbations in the locations of the input landmarks to introduce robustness to landmark uncertainty.

### 3.2 Cardiac Landmarks

The shape model is conditioned on nine anatomic landmarks: the apex, four on Mitral valve annulus (MV) and four at mid-level (Mid) of the LV (Fig 2). For training, the input landmarks were extracted automatically from parametric GT LV meshes, where each point corresponded roughly to the same anatomic location. During inference, these landmarks can either be provided externally (such as by a user) for a semi-automatic workflow, or automatically detected on the modality. For the automatic pipeline, we train deep reinforcement learning (DRL) based landmark detectors[32] on all three modalities. Any missed landmarks are estimated from the mean mesh (Appendix 1.3).

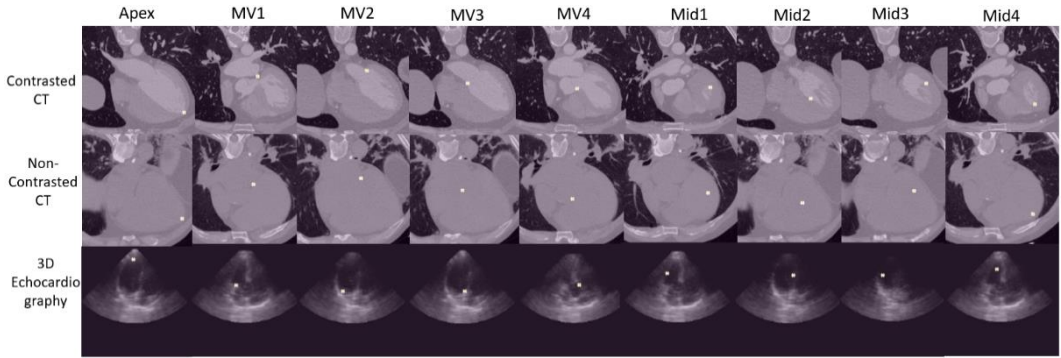

Fig. 2 Input cardiac landmarks. Each landmark is shown on one case from each modality. MV: Mitral Valve.

### 3.3 Local Refinement

Using landmarks as the input to the shape model can result in overlooking of fine-grained details available in the image. For the contrast CT and echocardiography data, we thus add a local refinement step that takes the input image and initial output of the shape model as two separate channels, and outputs a final segmentation. This was trained with GT segmentations of the respective datasets. For both modalities, the refinement is done using 64x64x64 patches, and uses a lightweight 3D UNet model with 3 downsampling blocks, each with a Conv3D, BatchNorm and Leaky RelU layers. The patches are selected to lie on the edges of the initial segmentation. The refinement is not performed on the non-contrast CT due to the lack of local image information.

### 3.4 Study Design

We train the shape model on LV landmarks and ground truth segmentations from contrasted CT scans, and evaluate on contrasted CT, non-contrasted CT and 3D TTE echocardiograms (3DE). It is to be noted that the same model trained on contrasted CT is used for all use cases without any further retraining. We compare the following:

a) Baseline: A 3D UNet is trained on each dataset as the baseline. The 3D UNet has 5 downsampling blocks, with increasing number of channels: 32, 64, 128, 256, 256). Each downsampling block has Conv3D, BatchNorm and Leaky RelU layers. Upsampling is done through nearest neighbor interpolation. Each network is trained for 300 epochs and best epoch is chosen using the validation dataset. We use Jaccard loss function and ADAM optimizer, with a learning rate of 0.001. All data is resampled to 1 mm resolution before training. Training is done patch-wise, due to GPU memory limitations. A patch size of 128x128x32 is used for the CT modalities, with 128 in-plane patch size and 32 in the axial direction. Since the echocardiography data is isotropic in all direction, a patch size of 128x128x128 is used. Random rotations and translations in 3D are used as data augmentation.

And DCSM based experiments:

b) Fully Automatic DCSM: The input landmarks are automatically found. Patients with missing landmarks are excluded.
c) Fully Automatic DCSM w. Estimation: Same as (b), but missing landmarks are estimated from the mean model and included.
d) Fully Automatic DCSM w. Estimation + Refinement: Same as (c), and with the refinement step (all except non-contrast CT).

e) Semi-automatic DCSM: We assume minimal error in the input landmarks, representing the upper limit of achievable accuracies. The input landmarks are extracted from the parametric GT meshes.

Reported metrics are Dice score as percentage, average surface distance, Hausdrorff distance and absolute and relative volume errors.

## 4 Data

**Contrast-enhanced CT.** The data consists of gated CCTA scans from 1474 patients. The data was randomly split on the patient level, into training (1197), validation (149) and testing (128) sets, which was maintained for all models trained on the data. The data was anonymized according to HIPAA standards. The datasets were acquired using Siemens SOMATOM CT scanners (Force, Definition Flash, Definition AS+). GT was obtained by running a previously validated segmentation algorithm[33] and visual review of every case to ensure accuracy of contours. The GT meshes are parametrized by 545 vertices, and including mitral annulus and outflow track.

**Non-contrast CT.** The data consists of gated, non-contrast scans acquired for calcium scoring for the same cohort as before. Since annotating chambers in non-contrast scans is a difficult task with high inter user variability, the GT contours were obtained from the corresponding contrasted scans for each patient. GT contours registered to the non-contrast scans using header information from the scans, followed by visual review to ensure accuracy. Additionally, we only include patients with contrasted and non-contrasted data acquired within 20% of the cardiac phase from each other, to minimize inaccuracies due to phase differences. Following these criteria, we obtain a cohort of 715 patients with non-contrast data and GT, which were divided into 595, 58 and 62 patients for training, validation and testing respectively. The training and validation datasets were only used for developing the landmark detection model. The patient splits were chosen to align with the splits in contrasted CT data.

**3D TTE Echocardiograms.** The data consists of TTE 3DE images from an independent cohort of 1287 patients from multiple centers. The data was randomly split patient wise into training (1005), validation (106) and testing (176) sets. The data was collected using the Siemens Healthineers SC2000, and ED and ES frames were selected for annotation. Each frame is a grayscale 8-bit 3D image, with isotropic spatial resolution of 1mm, of size 256x256x256. GT segmentations for the left ventricle were created by trained annotators. The annotator aligned the multiplanar view to obtain the true long axis of the LV in the A3C and A2C view. Landmarks were placed on the mitral annulus and apex. A mean LV mesh was positioned and adjusted manually where needed.

## 5   Results

**Input landmarks.** We obtain an average error of 4 mm, 7 mm and 8 mm for the contrasted CT, non-contrast CT and 3DE datasets respectively. The errors for each landmark and number of missed landmarks in each dataset are given in Appendix 1.4. Apex and mitral valve landmarks have lower errors in general compared to the mid-level landmarks, as they are more anatomically distinct.

**Segmentation.** The results of the evaluation for each dataset are given in Table 1-3. Inter-user variability for each modality was evaluated on a sub-set of the data by having two expert annotators manually delineate the LV. In case of non-contrast CT, we observe the best results using the fully automatic DCSM method. For the contrasted CT and 3DE dataset, the automatic DCSM followed by refinement outperforms the baseline in all metrics. As seen in Fig 3, while the shape model gives reasonable outputs, the refinement step allows better attention to detail. In all modalities, semi-automatic DCSM shows good performance, despite being developed on only contrasted CT data.

**Table 1.** Results on testing set for contrasted CT. All metrics are expressed as mean ± SD

| Model | #Patients | Dice (%) | Average Surface Distance (mm) | Hausdorff Distance (mm) | Volume errors (mL) | Relative Volume Errors (%) |
|---|---|---|---|---|---|---|
| Inter-user variability | 10 | 89.56 ± 4.0 | 1.98 ± 0.6 | 17.61 ± 3.6 | 13.26 ± 12.9 | 8.37 ± 8.1 |
| Baseline | 128 | 94.97 ± 9.6 | 0.86 ± 0.9 | 14.29 ± 16.4 | 4.75 ± 12.2 | 3.53 ± 9.7 |
| DCSM - Automatic | 125 | 88.12 ± 3.3 | 2.19 ± 0.5 | 17.73 ± 2.8 | 9.63 ± 13.0 | 5.57 ± 11.0 |
| DCSM - Automatic – w. Estimation | 128 | 87.90 ± 3.6 | 2.22 ± 0.5 | 17.83 ± 2.9 | 9.74 ± 13.3 | 5.62 ± 11.1 |
| DCSM - Automatic – w. Estimation + Refinement | 128 | **96.25 ± 1.7** | **0.69 ± 0.2** | **10.67 ± 6.5** | **1.18 ± 4.2** | **1.84 ± 3.9** |
| DCSM - Semi-automatic | 128 | 93.28 ± 2.2 | 1.28 ± 0.2 | 16.43 ± 3.1 | 2.53 ± 3.3 | 2.14 ± 3.3 |

**Table 2.** Results on testing set for non-contrasted CT. All metrics are expressed as mean ± SD

| Model | #Patients | Dice (%) | Average Surface Distance (mm) | Hausdorff Distance (mm) | Volume Errors (mL) | Relative Volume Errors (%) |
|---|---|---|---|---|---|---|
| Inter-user Variability | 12 | 77.26 ± 7.6 | 3.45 ± 1.4 | 12.75 ± 4.3 | -22.58 ± 23.9 | -29.8 ± 28.9 |
| Baseline | 62 | 80.76 ± 7.8 | 3.98 ± 2.2 | 39.49 ± 39.5 | -7.30 ± 27.7 | -9.43 ± 27.6 |
| DCSM - Automatic | 61 | **81.02 ± 7.5** | **3.28 ± 1.2** | **10.56 ± 3.3** | 2.23 ± 27.1 | -2.10 ± 28.2 |
| DCSM - Automatic – w. Estimation | 62 | 80.86 ± 7.6 | 3.31 ± 1.2 | 10.60 ± 3.2 | **2.20 ± 26.9** | **-2.06 ± 27.9** |
| DCSM - Semi-automatic | 62 | 94.56 ± 2.1 | 0.92 ± 0.2 | 4.18 ± 1.0 | -0.50 ± 3.3 | -0.44 ± 3.7 |

**Table 3.** Results on testing set for 3DE. All metrics are expressed as mean ± SD

| Model | #Patients | Dice (%) | Average Surface | Hausdorff Dis- | Volume Errors | Relative Volume |
|---|---|---|---|---|---|---|

| | | Distance (mm) | tance (mm) | (mL) | Errors (%) |
|---|---|---|---|---|---|
| Inter-user variability | 150 | 87.76 ± 5.7 | 2.44 ± 1.1 | 10.10 ± 4.3 | -4.20 ± 26.7 | -3.63 ± 16.3 |
| Baseline | 176 | 86.02 ± 5.4 | 2.49 ± 1.0 | 14.61 ± 11.1 | 2.51 ± 22.3 | 1.13 ± 20.4 |
| DCSM - Automatic | 161 | 83.46 ± 6.0 | 2.81 ± 1.1 | 9.81 ± 3.0 | 14.67 ± 20.8 | 11.10 ± 16.7 |
| DCSM - Automatic – w. Estimation | 176 | 82.80 ± 6.5 | 2.93 ± 1.2 | 10.12 ± 3.2 | 14.18 ± 21.6 | 10.82 ± 17.9 |
| DCSM - Automatic – w. Estimation + Refinement | 176 | **87.42 ± 5.9** | **2.20 ± 1.0** | **9.62 ± 6.5** | **2.09 ± 19.8** | **0.91 ± 17.2** |
| DCSM - Semi-automatic | 176 | 92.04 ± 1.4 | 1.39 ± 0.3 | 5.60 ± 1.1 | -6.5 ± 6.5 | -5.59 ± 4.8 |

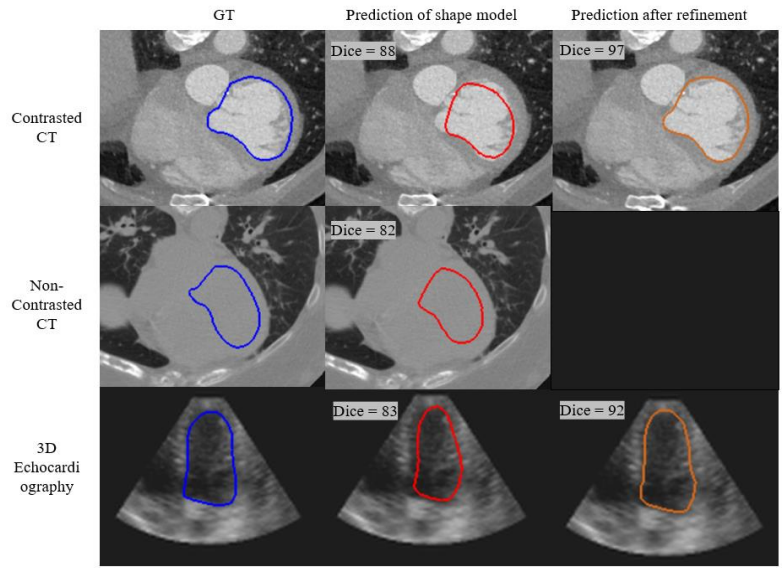

**Fig. 3.** GT and detections for some representative cases. Note: Refinement is not done for non-contrasted CT

## 6 Discussion

In this work, we have presented Deep Conditional Shape models (DCSMs) for medical image segmentation. The DCSM uses continuous, implicit neural representations at its core to represent the anatomical shape. The shape model is modality agnostic and can be trained with any available dataset and re-used for other modalities without any further re-training. In this work we train the shape model only on contrasted CT data, due to its large availability and ease of annotation, and apply it on non-contrast CT and 3DE data. Using the shape model improves the largest segmentation errors, as observed from significant improvements in the Hausdorff distances.

The shape model uses anatomical landmarks to constrain the position of the anatomy in 3D space. Thus, the trained shape model can be applied to a new modality given the landmarks, which can be obtained from a user (semi-automatic), or fully automatically using trained landmark detectors. The semi-automatic method obtains high performance with average dice greater than 92% and average distance error of less than 1.5 mm for all modalities. In this scenario, the DCSM is only trained on contrasted CT data, yet shows very high accuracy in the other unseen modalities. This is representative of scenarios where a user gives the landmarks or corrects the automatically detected landmarks with high confidence. This is also useful in multi-modality scenarios where landmarks from a high-fidelity modality (e.g.: contrasted CT) can be transformed to a lower fidelity one (e.g.: non-contrasted CT) using scanner information.

We also train DRL based networks to detect the landmarks automatically since manually locating the landmarks might not be feasible or convenient in all scenarios. This only requires the annotation of landmarks on the target modality, as opposed to full segmentation masks, and gives reasonable masks. It is observed that the shape model is sensitive to errors in the input landmarks, which can be mitigated to some extent using input perturbations during the model training. However, the automatic DCSM method shows lower accuracies than the semi-automatic one. The DRL landmarks might not produce landmarks in cases with low confidence, and these missing landmarks were estimated using the mean shape model. This estimation introduces further errors, as shown in the results. More sophisticated estimation methods could thus improve the accuracy in cases with missing landmarks.

A limitation of the shape model is that the input landmarks provide sparse information, as opposed to the fine-grained details provided by an image. To remedy this, we add a refinement step for modalities with image information such as the contrast CT and 3D echocardiography. The automatic DCSM with refinement outperforms the baseline in all metrics in contrasted CT and 3DE. The refinement step is not done for non-contrast CT due to the lack of image information. In that case, the automatic DCSMs outperform the baseline without the refinement. Using the shape model gives a clear advantage in case of modalities with very little image information, such as the non-contrast CT. In other cases, the refinement step allows for local refinement of details based on image information.

**Disclaimer:** The concepts and information presented in this paper/presentation are based on research results that are not commercially available. Future commercial availability cannot be guaranteed.

# 7 Appendix

## 7.1 Effect of number of input landmarks on accuracy

The impact of the number of input landmarks on output accuracy was studied on a separate test set of 50 patients. We consider the performance of the DCSM in the following input scenarios, arranged in the order of decreasing number of input points. Fourier mapping was not used here.

a) Complete mesh: This forms the upper limit of the performance of the shape model. Each mesh contains 545 points.

Sub-sampled scenarios:

b) Every-4: Sub-sample every 4th point of the full mesh,

c) Every 8: Sub-sample every 8th point of the full mesh,

Anatomically constrained reference points:

d) 9 landmark points: 1 point at Apex, 4 on Mitral Annulus, 4 at Mid level of the LV

e) 5 landmark points: 1 point at Apex, 4 on Mitral Annulus

As observed from Table 1, denser sampling of the points leads to better performance. The configuration with nine landmark points was chosen to balance accuracy, convenience, and interpretability.

| Experiments | Dice | Average Dist. | Hausdorff Dist. |
|---|---|---|---|
| All | 91.05 ± 1.7 | 1.55 ± 0.2 | 6.72 ± 1.5 |
| Every-4 | 89.26 ± 2.1 | 1.86 ± 0.3 | 6.73 ± 1.1 |
| Every-8 | 91.72 ± 1.8 | 1.47 ± 0.3 | 7.82 ± 1.9 |
| 9 landmark points | 87.00 ± 1.6 | 2.29 ± 0.2 | 7.84 ± 1.2 |
| 5 landmark points | 85.20 ± 2.1 | 2.62 ± 0.3 | 9.2 ± 1.5 |

**Table 1. Accuracies with different configurations of input landmarks**

### 7.2 Effect of Fourier mapping

The input points to the shape model are transformed with Fourier feature based mapping. We compare the results without the mapping (None), and with two types of mapping (Basic, Gaussian) [30] on a separate test set of 50 patients. All 545 points of the mesh were used as input. Both types of Fourier mapping give smoother training curves, and small but consistent improvements in the metrics, with Gaussian mapping giving the best results (Table 2).

**Table 2. Metrics with and without Fourier Mapping.**

| Experiments | Dice | Average Dist. | Hausdorff Dist. |
|---|---|---|---|
| None | 91.05 ± 1.7 | 1.55 ± 0.2 | 6.72 ± 1.5 |
| Basic | 91.83 ± 1.2 | 1.43 ± 0.18 | 6.67 ± 1.5 |
| Gaussian | **92.17 ± 1.3** | **1.36 ± 0.18** | **6.01 ± 1.1** |

### 7.3 Estimation of missing landmarks

Landmarks not detected by the DRL pipeline were estimated from the mean mesh. Let $P = \{p_i\}_{i=1}^{N} \in \mathbb{R}^3$ be the set of required input points. Let the set of detected points be $P_m = \{p_i\}_{i=1}^{m} \subset P$, where $m < N$. We find the affine transformation $H: P_m \to Q_m$ that transforms the detected set of points to the corresponding points in the mean mesh $Q$. Then we replace the missing landmarks with the same from the mean mesh and project them back into the patient space using the inverse transformation $H_{inv}$. This method introduces an estimation error to the landmark locations, as seen in the drop in accuracy between Automatic DCSM and Automatic DCSM w. Missing.

### 7.4 Baseline networks

We train a 3D UNet on each modality. The 3D UNet has 5 downsampling blocks, with increasing number of channels: 32, 64, 128, 256, 256). Each downsampling block has Conv3D, BatchNorm and Leaky RelU layers. Upsampling is done through nearest neighbor interpolation. Each network is trained for 300 epochs and best epoch is chosen using the validation dataset. We use Jaccard loss function and ADAM optimizer, with a learning rate of 0.001. All data is resampled to 1 mm resolution before training. Training is done patch-wise, due to GPU memory limitations. A patch size of 128x128x32 is used for the CT modalities, with 128 in-plane patch size and 32 in the axial direction. Since the echocardiography data is isotropic in all direction, a patch size of 128x128x128 is used. Random rotations and translations in 3D are used as data augmentation.

### 7.5 DRL Landmark Detection

Table 3. Detection **errors and rates for each landmark in each** modality.

|  | Contrast CT (128) | | Non-contrast CT (62) | | 3D Echocardiography (176) | |
| --- | --- | --- | --- | --- | --- | --- |
|  | Error (mm) | #Missing | Error (mm) | #Missing | Error (mm) | #Missing |
| Apex | 3.0 ± 1.5 | 1 | 7.0 ± 3.5 | 1 | 7.0 ± 3.8 | - |
| MV1 | 2.5 ± 1.4 | - | 5.6 ± 2.7 | - | 6.6 ± 5.0 | - |
| MV2 | 3.1 ± 1.9 | - | 6.1 ± 3.7 | - | 6.4 ± 3.9 | 1 |
| MV3 | 2.8 ± 1.9 | - | 6.1 ± 3.5 | - | 5.3 ± 3.4 | - |
| MV4 | 2.9 ± 2.3 | - | 5.1 ± 2.2 | - | 6.8 ± 6.8 | 2 |
| Mid1 | 5.7 ± 3.2 | - | 8.6 ± 4.3 | - | 11.1 ± 7.3 | 3 |
| Mid2 | 4.9 ± 3.1 | - | 7.8 ± 3.8 | - | 8.8 ± 5.3 | 3 |
| Mid3 | 5.5 ± 3.4 | 1 | 9.0 ± 3.8 | - | 9.7 ± 6.4 | 6 |
| Mid4 | 6.3 ± 4.2 | 1 | 8.6 ± 4.2 | - | 10.1 ± 6.1 | 3 |